\documentclass[a4paper,11pt]{article}
\pdfoutput=1 

\usepackage{jcappub} 

\usepackage[T1]{fontenc} 

\title{The dependence of halo bias on the protohalo shape alignment with the initial tidal field}

\author[a,1]{Jounghun Lee \note{Corresponding author.}}
\author[a,b]{Jun-Sung Moon}

\affiliation[a]{Department of Physics and Astronomy, Seoul National University, \\
Kwanak-ro 1, Kwanak-gu, Seoul 08826, Republic of Korea}
\affiliation[b]{Research Institute of Basic Sciences, Seoul National University,  \\
Kwanak-ro 1, Kwanak-gu, Seoul 08826, Republic of Korea}

\emailAdd{cosmos.hun@gmail.com}
\emailAdd{jsmoon.astro@gmail.com}

\abstract{We present a numerical evidence supporting the primordial origin of secondary halo bias even on the galactic mass scale.   
Analyzing the data from the TNG 300-1 simulations, we investigate the dependence of halo bias on the degree of misalignment between the protohalo inertia and 
initial tidal tensors, $\tau$, measured at redshift, $z_{i}=127$. From the TNG 300-1 galactic halos in logarithmic mass range of $10.5< m\equiv \log[M/(h^{-1}M_{\odot})]\le 13$ 
identified at $z=0,\ 0.5$ and $1$, a clear signal of $\tau$ bias is detected. For the case that $\tau$ is measured from the initial tidal field smoothed on the scale of 
$R_{f}/(h^{-1}\,{\rm Mpc})\lesssim 1$,  the halo $\tau$ bias is found to be very similar in its tendency and amplitude to the spin bias at all of the three redshifts, 
if the effects of backsplash halos are properly eliminated. 
For the case of $R_{f}/(h^{-1}\,{\rm Mpc})=2$, the $\tau$ bias at $z=1$ turns out to behave like the age bias, diminishing rapidly in the range of $m> 12$.  
At $z=0$ and $0.5$, however, the $\tau$ and age bias factors show large differences in their overall strengths, which is attributed to the dominant nonlinear effects 
that undermine the former but enhance the latter. 
Given these numerical results along with the previous finding~\cite{ML24} that $\tau$ shares a large amount of mutual information with the formation epochs and 
spin parameters of galactic halos, it is concluded that the origins of halo age and spin bias must be closely linked with the primordial factor, $\tau$, and that 
the difference in the tendency between the two bias factors on the galactic mass scale reflects the multi-scale influence of $\tau$ on the halo secondary properties.}
\begin{document}
\maketitle
\flushbottom

\section{Introduction}\label{sec:intro}

The dark matter halos are biased tracers of the underlying dark matter (DM) particles, as the former exhibits mass-dependent difference in their clustering 
strengths from the latter~\cite{kai84,MW96}. The halos with mass $M$ higher than the characteristic mass scale $M_{\ast}$ tend to cluster more strongly than the DM particles, 
while the opposite tendency is exhibited by the low-mass counterparts with $M\lesssim M_{\ast}$. Here, $M_{\ast}$ satisfies the condition of $\sigma(M_{\ast})=\delta_{c}$ 
with critical density contrast for collapse, $\delta_{c}$, and rms density fluctuation, $\sigma$~\cite{GG72}. 
This mass dependence of halo bias is predicted by the standard excursion set theory~\cite{PS74,bon-etal91,MW96}, which effectively describes the growths of initial density peaks 
till their gravitational collapse as random walking processes. 

It was, however, discovered by multiple N-body simulations that the halo bias depends not only on mass but also on secondary properties like formation epoch, concentration, 
substructure abundance, velocity anisotropy, potential depth, accretion rate, shape and spin parameter as well
~\cite{gao-etal05,wec-etal06,GW07,li-etal08,hah-etal09,fal-etal10,LP12,li-etal13,sun-etal16,laz-etal17,zomg1,sal-etal18,con-etal18,joh-etal19,sat-etal19,mon-etal20,MK20,wan-etal21,bal-etal23,MR23,com-etal24}. 
Among these secondary properties, what turned out to exhibit the strongest bias signal is the halo age for which the formation epoch and concentration parameter often stand 
proxy~\cite{gao-etal05,GW07,li-etal13,sal-etal18,joh-etal19,sat-etal19}. In the lowest-mass section ($M\ll M_{\ast}$) is found a positive age bias trend that the older halos cluster more 
strongly. While in the highest-mass section ($M\gg M_{\ast}$) is detected the opposite tendency that the younger halos cluster more strongly, i.e., negative age bias. 
The inversion from negative to positive age bias trend was witnessed around the characteristic mass scale $M_{\ast}$~\cite{sat-etal19}.  Regarding the other secondary properties, 
their bias trends were shown to be similar to that of the halo ages, except for the spin bias, which inverts at a much lower mass threshold 
$M\ll M_{\ast}$~\cite{sal-etal18,sat-etal19,joh-etal19,splash_spin}.

Ever since the first discovery of this phenomenon~\cite{gao-etal05}, much theoretical effort was made to physically understand why the DM halos with different 
secondary properties cluster differently. Although no standard theory has yet to be fully established, it was conventionally accepted that the secondary bias of low-mass halos 
differs in its origin from that of high-mass counterparts. In the highest-mass section ($M\gg M_{\ast}$), the secondary halo bias was shown by ref.~\cite{dal-etal08} to have a primordial 
origin.  Incorporating the peak curvature dependence of initial density correlations into the excursion set theory, ref.~\cite{dal-etal08} analytically proved that the Lagrangian halo bias 
factor in the asymptotic limit of $M\gg M_{\ast}$ is a function of the curvature of initial density peaks, on which the halo formation epochs depend. 

Whereas, in the lowest-mass section ($M\ll M_{\ast}$),  the secondary bias was often attributed to the nonlinear physical processes during their assembly histories. 
Ref.~\cite{MK20} scrupulously examined various theories suggested by the previous works for the origin of low-mass age bias and found that the following three 
mechanisms should make the most significant contributions: first, the presence of backsplash halos in the vicinity of their former higher-mass 
hosts~\cite{wan-etal09,li-etal13,splash_spin};  second, the effects of large-scale web environments which can differentiate the formation epochs of 
halos at fixed mass by either facilitating or impeding mass accretion~\cite{zomg1}; third, the shock heat generated during the collapse of largest-scale structures, 
which can slow down the mass accretion into the filament and sheet halos~\cite{wan-etal07,dal-etal08}.  

In the analysis of ref.~\cite{MK20}, however, it was also found that when the backsplash halos were excluded, the fraction of low-mass halos yielding positive age bias 
was quite low, which implied that not only the highest-mass halos but also the lower mass halos may have had negative age bias of primordial origin until they were vulnerably 
exposed to the aforementioned physical mechanisms. 
This idea, {\it the primordial origin of secondary halo bias} on the lower-mass scale of $M\lesssim M_{\ast}$, was also supported by the result of ref.~\cite{wan-etal21} that the 
secondary halo properties are more strongly correlated with the linear density field rather than with the evolved density field.  According to ref.~\cite{wan-etal21}, the secondary 
halo bias should be a collateral byproduct of the multi-scale competition between the external and internal correlations of secondary halo properties with the linear density field, 
where the external and internal correlations are measured on the scales larger and smaller than the Lagrangian halo radius, respectively. 
In these previous works, however, it was still inconclusive which primordial factor actually drives the halo bias to acquire dependence on the secondary properties even on the 
mass scale of $M\lesssim M_{\ast}$, and what causes the age and spin bias to behave differently on this mass scale~\cite{sat-etal19,joh-etal19,splash_spin}, if both of them share the same 
origin~\citep{LP12,joh-etal19}. 

For the verification of the primordial origin of the secondary bias of lower mass halos,  it is necessary to find out a primordial factor on which the lower-mass halo bias 
exhibits the same strength of dependence as (or at least similar to) on the other secondary properties. In addition, a proper explanation for the difference between the age and 
spin bias in terms of this singled out primordial factor is also required. In light of our prior work where the ages and spins of galactic halos were found to share a large amount of 
mutual information with the degree of misalignments, $\tau$, between the principal axes of protohalo inertia and initial tidal tensors, we put forth a conjecture that $\tau$ may be the 
primordial factor, and attempt to verify this conjecture by exploring the halo $\tau$ bias.  As in ref.~\cite{MK20}, we will focus on the galactic mass scale, 
$10.5< m\equiv \log\left[M/(h^{-1}\,M_{\odot})\right]\le 13$, where the existence of the secondary bias of primordial origin remains a crux, and  the trend of spin bias significantly 
deviates from that of age bias~\cite{sal-etal18,sat-etal19,joh-etal19,splash_spin}.
 
The outlines of the upcoming sections are as follows. 
In section~\ref{sec:review},  the halo age and spin bias factors are reviewed, and their mutual differences on the galactic mass scales are reinspected. 
In section~\ref{sec:tau_bias}, the dependence of the halo bias on the primordial factor $\tau$ and the variation of its strength with mass and redshift 
are investigated, and compared with those of the halo age and spin bias. 
In section~\ref{sec:con},  the key results are summarized and their physical implications are discussed. 

\section{Detection of halo $\tau$ bias and its implication on the secondary bias}

\subsection{A brief review of halo age and spin bias trends}\label{sec:review}

\begin{figure}[tbp]
\centering 
\includegraphics[width=0.85\textwidth=0 380 0 200]{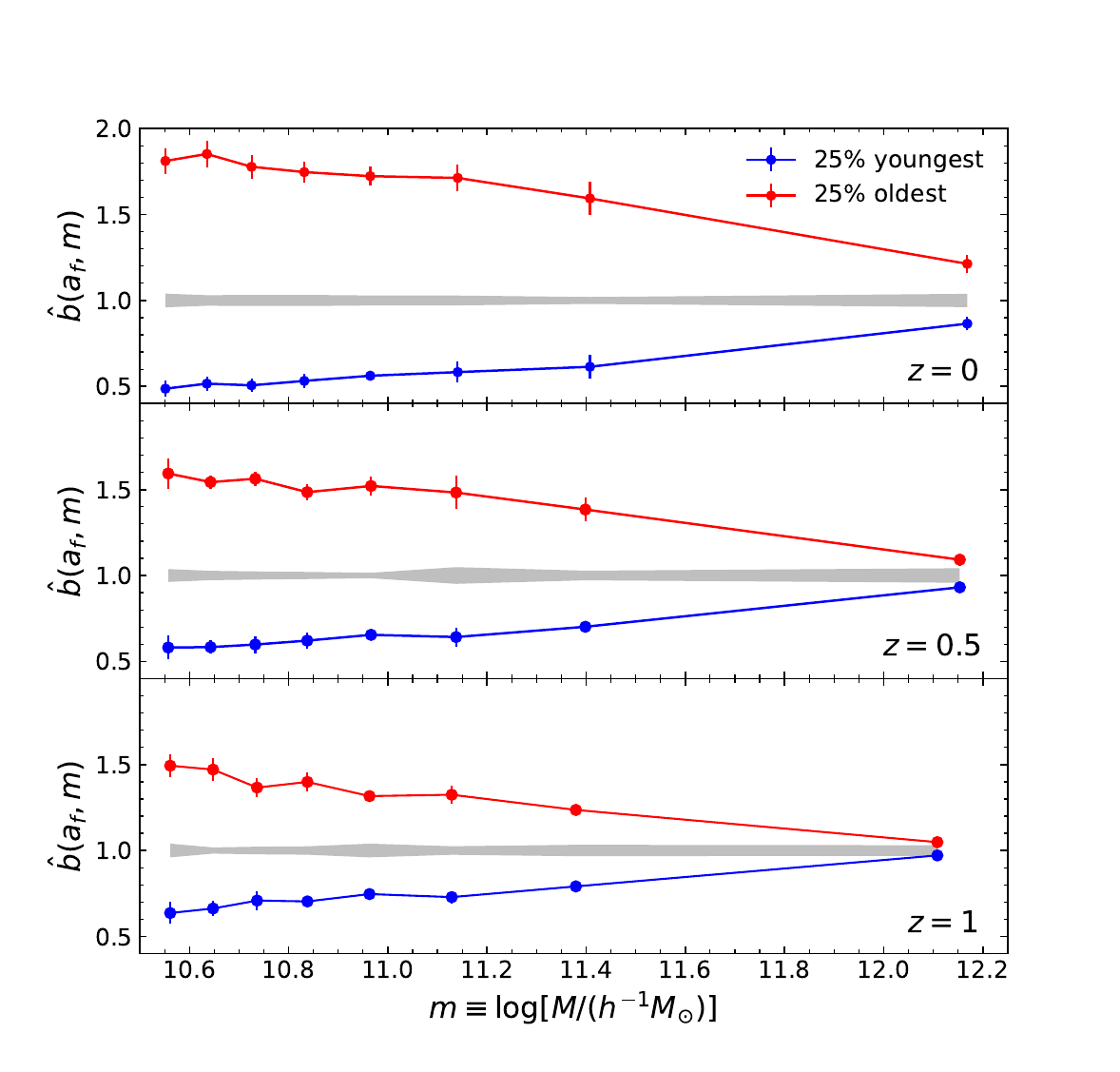}
\caption{\label{fig:age_bias} Relative age bias, $\hat{b}(a_{f},m)$, from the $8$ $m$-selected subsamples of the galactic halos belonging to the top (red filled circles) and 
bottom (blue filled circles) quartiles of their ages versus the mean value of $m$ averaged over each $m$-selected subsample. at three different redshifts.  
In each panel, the horizontal the grey filled area represents the $1\sigma_{b}$ Jackknife scatter around the relative halo bias 
$\hat{b}(m)$ from the whole population of the galactic halos in the logarithmic mass range of $10.5\le m\le 13$.}
\end{figure}
\begin{figure}[tbp]
\centering 
\includegraphics[width=0.85\textwidth=0 380 0 200]{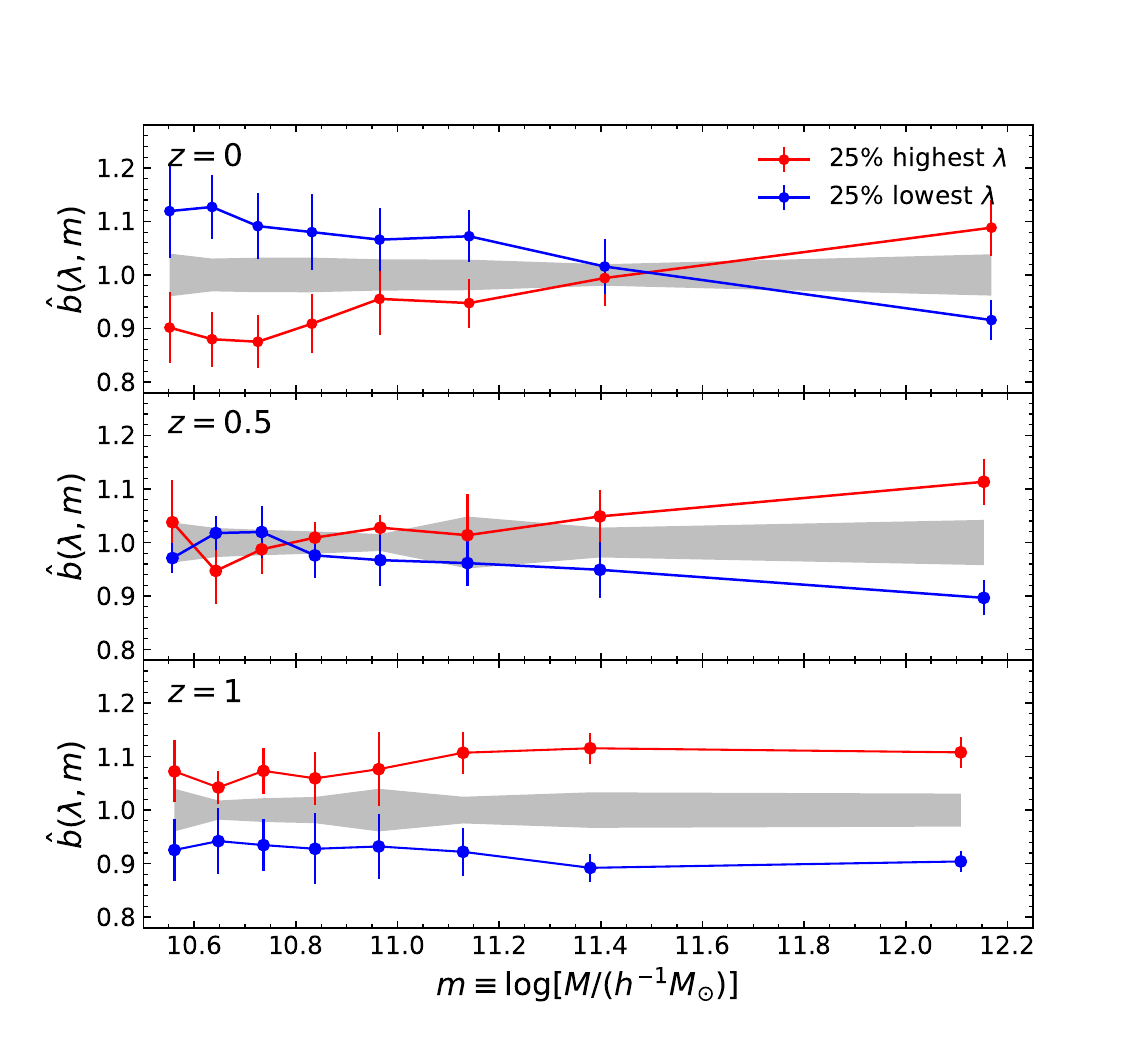}
\caption{\label{fig:spin_bias} Relative spin bias from the top (red filled circles) and bottom (blue filled circles) quartiles of the halo spin parameters. $\lambda$. }
\end{figure}
\begin{figure}[tbp]
\centering 
\includegraphics[height=410 pt,width=395 pt]{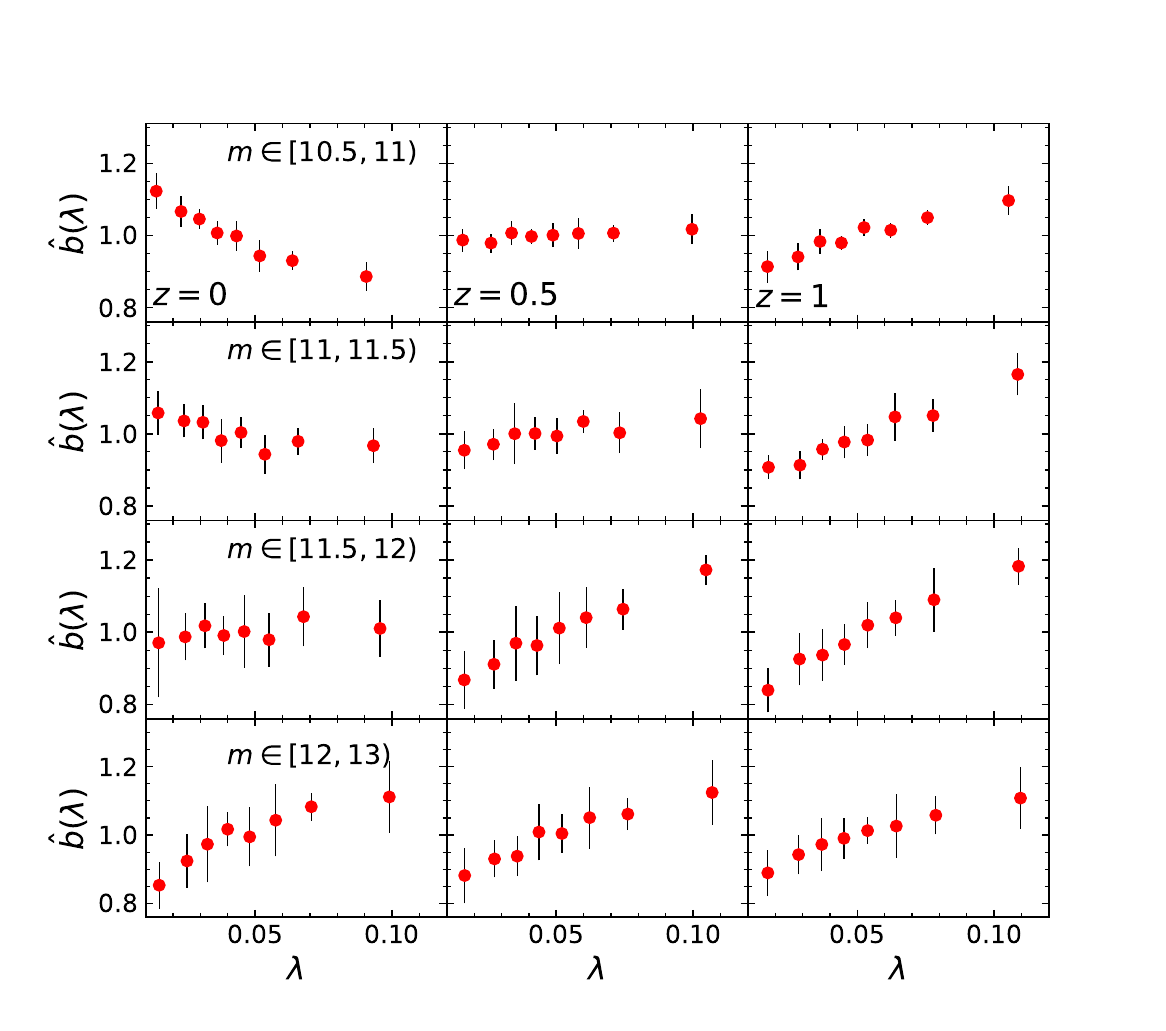}
\caption{\label{fig:spin_bias2} Relative spin in the whole range of $\lambda$ (red filled circles) from the galactic halos belonging 
to four different $m$-bins at $z=0$, $0.5$ and $1$.}
\end{figure}

To numerically explore if the secondary halo bias has a primordial origin even on the galactic mass scale of $10.5< m \le  13$, we utilize the data from the 300 Mpc volume run 
(TNG300-1) of the IllustrisTNG suite of cosmological gravo-magneto-hydrodynamical simulations~\citep{tng1,tng2,tng3,tng4,tng5, tng6}. 
The TNG 300-1 run was conducted on a box of volume $V_{\rm tot}\equiv (205)^{3}\,h^{-3}{\rm Mpc}^{3}$ 
with the Planck initial condition~\cite{planck15} to track the dynamical evolution of $2\times 2500^{3}$ particles from the initial state $z_{i}=127$ down to the present epoch $z=0$. 
One half of the particles in the TNG 300-1 are baryonic cells with individual mass of $1.1\times 10^{7}\,M_{\odot}$, while the other half are collisionless DM 
particles each of which is as massive as $5.9\times 10^{7}\,M_{\odot}$. 
The standard friends-of-friends (FoF) with the linkage length parameter of $0.2$ and SUBFIND algorithms~\cite{subfind} were employed to resolve the bound halos and their 
substructures, respectively, in the TNG 300-1 snapshots~\citep{tng3}. Selecting the most massive central subhalos of the FoF groups, we extract information on their comoving 
positions (${\bf x}$), total masses ($M$), formation epochs ($a_{f}$) as defined in ref.~\cite{LC93,LC94} and spin parameters ($\lambda$) as defined in ref.~\cite{bul-etal01} 
from the publicly released dataset of the TNG 300-1 simulations\footnote{https://www.tng-project.org/data/}. 
Throughout this paper, we will refer to the selected central subhalos as {\it galactic halos}, and consider only two secondary properties, $a_{f}$ and $\lambda$, of them. 

The Eulerian bias factor of dark halos, $b(m)$, at a given redshift is defined as $b(m)\equiv \delta_{h}(m)/\delta$ where $\delta_{h}(m)$ is the number density 
contrast of the halos with logarithmic masses in a differential interval of $[m,m+dm]$ within an overdense region of Eulerian volume $\Delta V$~\cite{MW96}, while $\delta$ 
is the matter density contrast within $\Delta V$. Applying the extended excursion set theory, ref.~\cite{MW96} evaluated $\delta_{h}(m)$ as 
\begin{equation}
\label{eqn:bias}
\delta_{h}(m) \equiv\frac{n(m | \delta>0)}{n(m)\Delta V} - 1\, , 
\end{equation}
where  $n(m)$ is the unconditional mass function of halos observed at $z$, while $n(m|\delta>0)$ is the conditional mass function of those halos embedded in $\Delta V$.  
In the asymptotic limit where $\Delta V$ is much larger than the volumes of individual halos with logarithmic mass $m$, $b(m)$ is called the Eulerian {\it linear} bias factor and 
shown to be equal to the Lagrangian bias plus unity~\cite{MW96}. Throughout this paper, we will focus on the Eulerian linear bias and its dependence on the halo 
secondary properties. 

The following flow chart describes how we determine the relative age and spin bias factors from the TNG 300-1 data in accordance with eq.(\ref{eqn:bias}):
\begin{enumerate}
\item
Divide the simulation volume, $V_{\rm tot}$, into $64^{3}$ grids of equal volume $\Delta V\equiv V_{\rm tot}/(64^{3})$. This choice of grid number is made to 
ensure that $b(m)$ is the {\it linear} bias factor. Given that the onset of nonlinear bias is witnessed on the scales below a typical cluster size $\sim 2\,h^{-1}$Mpc~\cite{nlbias}, 
the evaluation of a linear bias factor requires $\Delta V > \left(2\,h^{-1}{\rm Mpc}\right)^{3}$. The grid number, $64^{3}$, satisfies this condition and simultaneously 
allows us to avoid poor number statistics in the evaluation of secondary bias, which can be caused when  grid number is too small. 
\item
Construct the matter density contrast field, $\delta$, on the grids by applying the cloud-in-cell algorithm to the particle snapshot at each of the three redshifts, and
select the galactic halos in the logarithmic mass range of $10.5< m\equiv \ln [M/(h^{-1}\,M_{\odot})]\le 13$ to make a main sample. 
\item
Split the main sample into eight subsamples of equal size according to the values of $m$. 
Treat the galactic halos belonging to each $m$-selected subsample as particles, construct the halo number density contrast field, $\delta_{h}(m)$, with the help of the 
cloud-in-cell method. 
 \item
Examine the $\delta$ value at each of $64^{3}$ grids and select those grids at which $\delta \ge \sigma$ to minimize the shot noise. 
Take the ratio, $\delta_{h}(m)/\delta$, at each of the selected grids, and determine $b(m)$ as the ensemble average of this ratio over the selected grids 
for each of the $8$ $m$-selected subsamples. 
\item
Create eight jackknife resamples by dividing the selected grids points according to the grid locations, and compute the Jackknife errors 
from the eight resamples (one standard deviation scatter of the bias factor obtained from each resample from the original $b(m)$ obtained 
over $V_{\rm tot}$. 
\item
Repeating the above steps (2)-(5), but with only those youngest (oldest) galactic halos whose formation epochs belong to the top (bottom) quartile values of $a_{f}$, 
determine the halo age bias, $b(a_{f},m)$, and take its ratio to $b(m)$ to obtain the relative age bias, $\hat{b}(a_{f},m)\equiv b(a_{f},m)/b(m)$. 
\item 
Repeating the above steps (2)-(5), but with only those rapidly (slowly) spinning galactic halos whose spin parameters belong to the top (bottom) quartile 
of $\lambda$, determine the spin bias, $b(\lambda,m)$ and then take its ratio to $b(m)$ to obtain the relative spin bias, $\hat{b}(\lambda,m)\equiv b(\lambda,m)/b(m)$. 
\end{enumerate}

A total of $618777$, $637406$ and $644753$ central subhalos are selected at $z=0$, $0.5$ and $1$, respectively. 
Figure~\ref{fig:age_bias} plots $\hat{b}(a_{f},m)$ versus the mean values of $m$ of eight $m$-selected subsamples of the galactic halos 
with $a_{f}$ belonging to its top and bottom quartiles (blue and red filled circles, respectively). As can be seen, our results are quite consistent with what was found 
in the previous works~\cite{gao-etal05,wec-etal06,GW07,hah-etal09,sun-etal16,sal-etal18,joh-etal19,sat-etal19}, capturing two characteristic tendencies of the age bias: 
the older halos ($25\%$ lowest $a_{f}$) cluster more strongly than the younger ones ($25\%$ highest $a_{f}$)~\cite{sat-etal19} and this age dependence 
of the halo bias becomes weaker  at higher $z$~\cite{con-etal18}.
Figure~\ref{fig:spin_bias} plots the same as figure~\ref{fig:spin_bias} but from the top and bottom quartiles of $\lambda$, confirming the existence of spin bias. 
As can be seen, the inversion from the negative to positive spin bias is witnessed around $m\simeq 11.5$ only at $z=0$. Meanwhile, at higher $z$ the spin bias 
appears to become more significant as $m$ increases in the mass range considered, showing no inversion. 
The overall trends of $\hat{b}(a_{f},m)$ and $\hat{b}(\lambda,m)$ are all quite consistent with the previous 
findings~\cite{gao-etal05,wec-etal06,GW07,li-etal08,hah-etal09,li-etal13,sun-etal16,sal-etal18,joh-etal19,sat-etal19}. 

As can be seen, the spin bias is less significant than the age counterpart on the galactic mass scale. To see more closely how the halo bias changes with $\lambda$ at each redshift, 
we split the main sample into four subsamples corresponding to the $m$-ranges of $[10.5, 11)$, $[11,11.5)$, $[11.5,12)$ and $[12,13]$.  Dividing the range of $\lambda$ of 
the galactic halos belonging to each of the four $m$-selected subsample into eight bins, we compute $\hat{b}(\lambda)$ separately at each $\lambda$ interval via the above flow chart. 
Figure~\ref{fig:spin_bias2} shows $\hat{b}(\lambda)$ versus $\lambda$ from the four $m$-selected subsamples at the three redshifts, revealing that the inversion of 
$\hat{b}(\lambda)$ occurs not only in a $m$-dependent way but also in a $z$-dependent way. The galactic halos of mass $10.5\le m\le 11$ decreases (increases) with
 $\lambda$ at $z=0$ ($z=1$), while at $z=0.5$, they show little variation with $\lambda$. 
 
It is worth mentioning here that we estimate the secondary halo bias by taking the excursion set approach based on eq.~(\ref{eqn:bias}) rather than by employing the 
conventional methods based on the density correlations or power spectra~\cite{gao-etal05,GW07,fal-etal10,sal-etal18,joh-etal19,sat-etal19,MK20}.  
True as it is that the conventional methods are quite efficient for the estimation of secondary halo bias in a wide mass range, it suffers from an ambiguity in the choice of the range 
of separation distance $r$,  which may make it difficult to properly investigate the dependence of halo bias on a scale-dependent secondary property like $\tau$, the primary focus 
of the current analysis.  Nevertheless, our method turns out to reproduce quite well the overall trends of age and spin bias estimated by the conventional correlation function method 
in the galactic mass range. 

\subsection{Scale and redshift dependence of halo $\tau$-bias}\label{sec:tau_bias}

As mentioned in section~\ref{sec:intro}, it was found by our prior work~\cite{ML24} that the basic traits of galactic halos share a large mount of mutual information with 
a primordial factor, $\tau$,  defined as the degree of misalignment\footnote{Since the principal axes of a protohalo inertia tensor are aligned with it shape axes, 
it is equivalent to saying that $\tau$ describes how weakly the protohalo shape axes are anti-aligned with the principal axes of the initial tidal field.} between the principal 
axes of protohalo inertia and initial tidal tensors.  
For the case of massive galactic halos with $m\ge 11.5$, the primordial factor $\tau$ was found to contain larger amounts of mutual information about the formation epochs and 
spin parameters than the halo mass, local density and environmental shears (see figure 3 in ref.~\cite{ML24}).  Furthermore, the probability density functions of both 
of $\lambda$ and $\tau$ were also shown to be best approximated by the same Gamma distribution~\cite{ML24}. 
These prior results naturally lead us to expect the existence of $\tau$ bias and to speculate its possible links with the age and spin bias. 

In the following, we concisely describe the computational steps via which the relative $\tau$ bias factor, $\hat{b}(\tau,m)$, is measured:
\begin{enumerate}
\item
 Trace back the constituent particles of each galactic halo back to the earliest epoch of redshift $z_{i}=127$ to 
 find their Lagrangian positions, ${\bf q}\equiv (q_{i})$, and their center of mass ${\bf q}_{c}\equiv (q_{c,i})$, as well. 
 Then, compute the inertia tensor of a located protohalo, $I_{jk}({\bf q}_{c})$, as~\cite{ML24}
\begin{equation}
\label{eqn:inertia}
\hat{I}_{jk} \equiv \sum_{\mu=1}^{n_{p}}m_{\mu}(q_{\mu,j}-q_{c,j})(q_{\mu,k}-q_{c,k})\, ,
\end{equation}
where $m_{\nu}$ is the mass of $\nu$th particle located at the initial position ${\bf q}_{\mu}\equiv (q_{\mu,i})$, and $n_{p}$ is the number of the particles 
that constitute the protohalo. 
\item
Apply the cloud-in-cell method to the particle snapshot at redshift $z_{i}=127$ to determine the initial matter density contrast field, $\delta({\bf q})$, 
on $512^{3}$ grid points as in~\cite{ML24}. 
Find the Fourier space initial density contrast field, $\delta({\bf k})$, via a Fast Fourier transformation (FFT) of $\delta({\bf q})$ 
with the Fourier space wave vector of ${\bf k}$.
\item 
Compute the Fourier space initial tidal field convolved by a Gaussian kernel with scale radius $R_{f}$, 
as  $\tilde{T}_{ij}({\bf k})\equiv k_{i}k_{j}\tilde{\delta}({\bf k})\exp(-\vert{\bf k}\vert^{2}R^{2}_{f}/2)/\vert{\bf k}\vert^{2}$. 
Then, perform an inverse FFT of it to obtain the real-space initial tidal field $[{T}_{ij}({\bf q})]$. 
At the grid point in which ${\bf q}_{c}$ falls, diagonalize $T_{ij}({\bf q}_{c})$ to find its three orthonormal eigenvectors, $\{{\bf e}_{1},\ {\bf e}_{2},\ {\bf e}_{3}\}$. 
\item
Construct a rotation matrix, ${\bf R}$, whose three columns correspond to $\{{\bf e}_{1},\ {\bf e}_{2},\ {\bf e}_{3}\}$, and 
perform a similarity transformation of the protohalo inertia tensor as $I^{\prime}_{jk}=R^{-1}_{jl}I_{ls}R_{sk}$. 
\item 
For each galactic halo, compute the degree of misalignment, $\tau(r_{f})$, between the principal axes of the protohalo inertia and initial tidal tensors smoothed 
on the scale of $r_{f}\equiv R_{f}/(h^{-1}{\rm Mpc})$ as~\cite{ML24}
\begin{equation}
\tau(r_{f}) \equiv \left[\frac{(I^{\prime}_{12})^{2}+(I^{\prime}_{23})^{2}+(I^{\prime}_{31})^{2}}{(I^{\prime}_{11})^{2}+(I^{\prime}_{22})^{2}+(I^{\prime}_{33})^{2}}\right]^{1/2}\, . 
\end{equation}
\item
Follow the flow chart described in section~\ref{sec:review} to obtain the relative $\tau$ bias factor for three different cases of 
$r_{f}=0.5$, $1$ and $2$, denoted by $\hat{b}(\tau_{0.5},m)$, $\hat{b}(\tau_{1},m)$ and $\hat{b}(\tau_{2},m)$, at each redshift.
\end{enumerate}

\begin{figure}[tbp]
\centering 
\includegraphics[height=410 pt,width=395 pt]{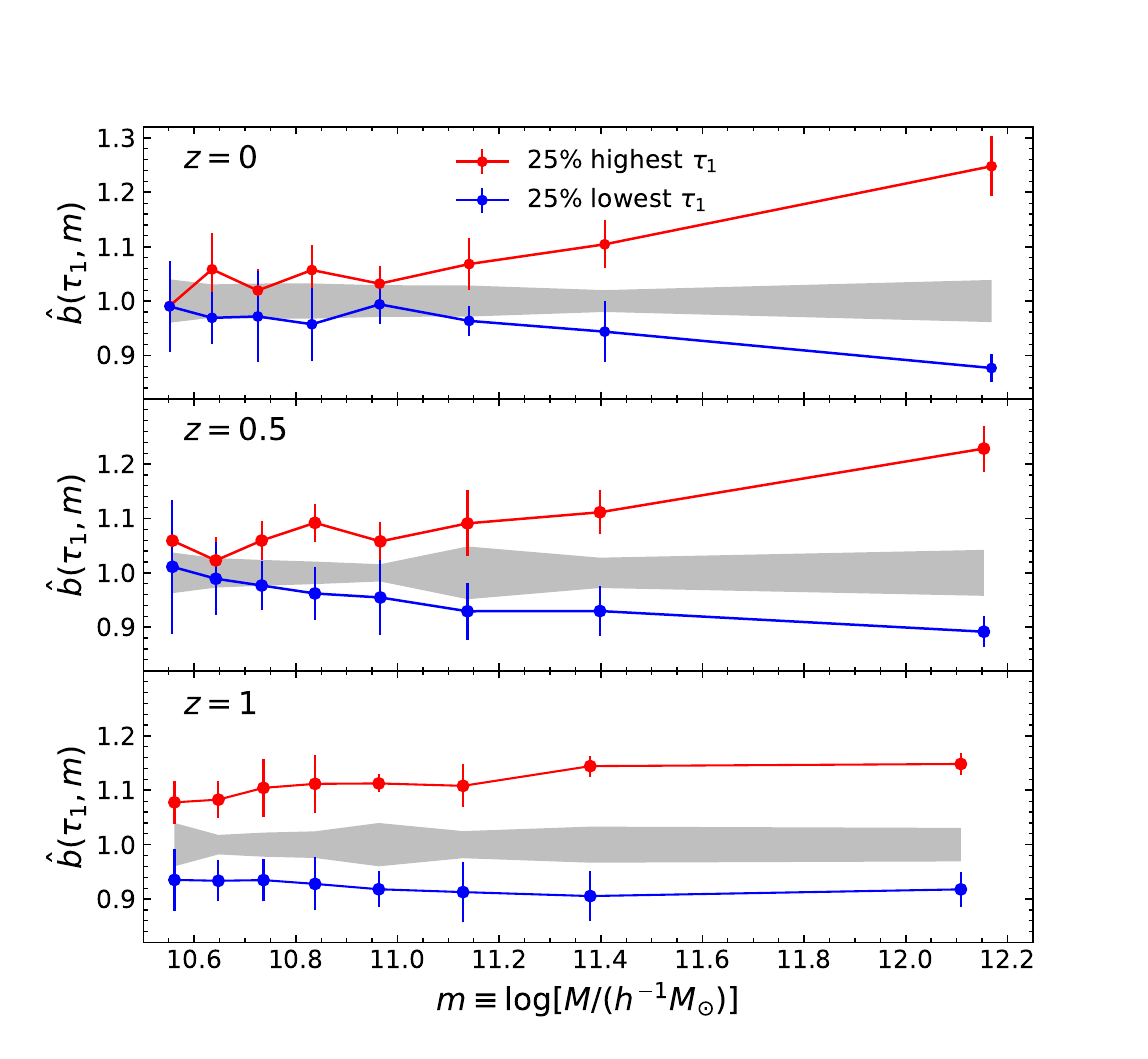}
\caption{\label{fig:tau1_bias} Relative $\tau_{1}$ bias from the galactic halos whose primordial $\tau_{1}$ values belong to the 
top (red filled circles) and bottom (blue filled circles) quartiles, where $\tau_{1}$ quantifies how weakly the triaxial proto-halos shapes 
are aligned with the principal axes of the initial tidal field smoothed on the scale of $R_{f}/(h^{-1}{\rm Mpc})=1$ (see eq.(\ref{eqn:inertia})). 
Note a robust similarity between $\hat{b}(\tau_{1},m)$ and $\hat{b}(\lambda,m)$ (see figure~\ref{fig:spin_bias}) at $z=1$.}
\end{figure}
\begin{figure}[tbp]
\centering 
\includegraphics[height=410 pt,width=395 pt]{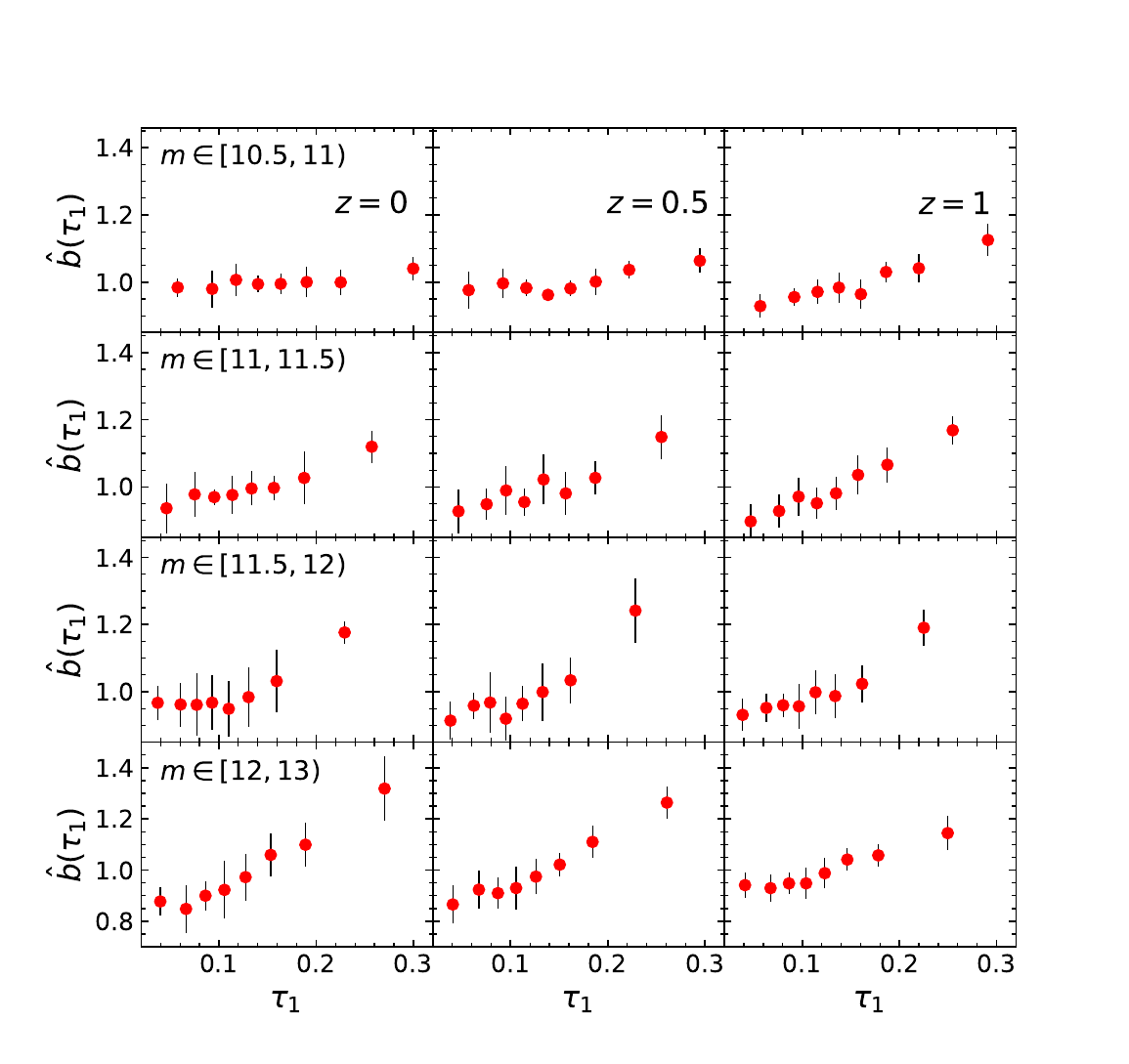}
\caption{\label{fig:tau1_bias2} Relative $\tau_{1}$ in the whole range of $\tau_{1}$ (red filled circles) from the galactic halos belonging 
to four different $m$-bins at $z=0$, $0.5$ and $1$.}
\end{figure}

\begin{figure}[tbp]
\centering 
\includegraphics[height=410 pt,width=395 pt]{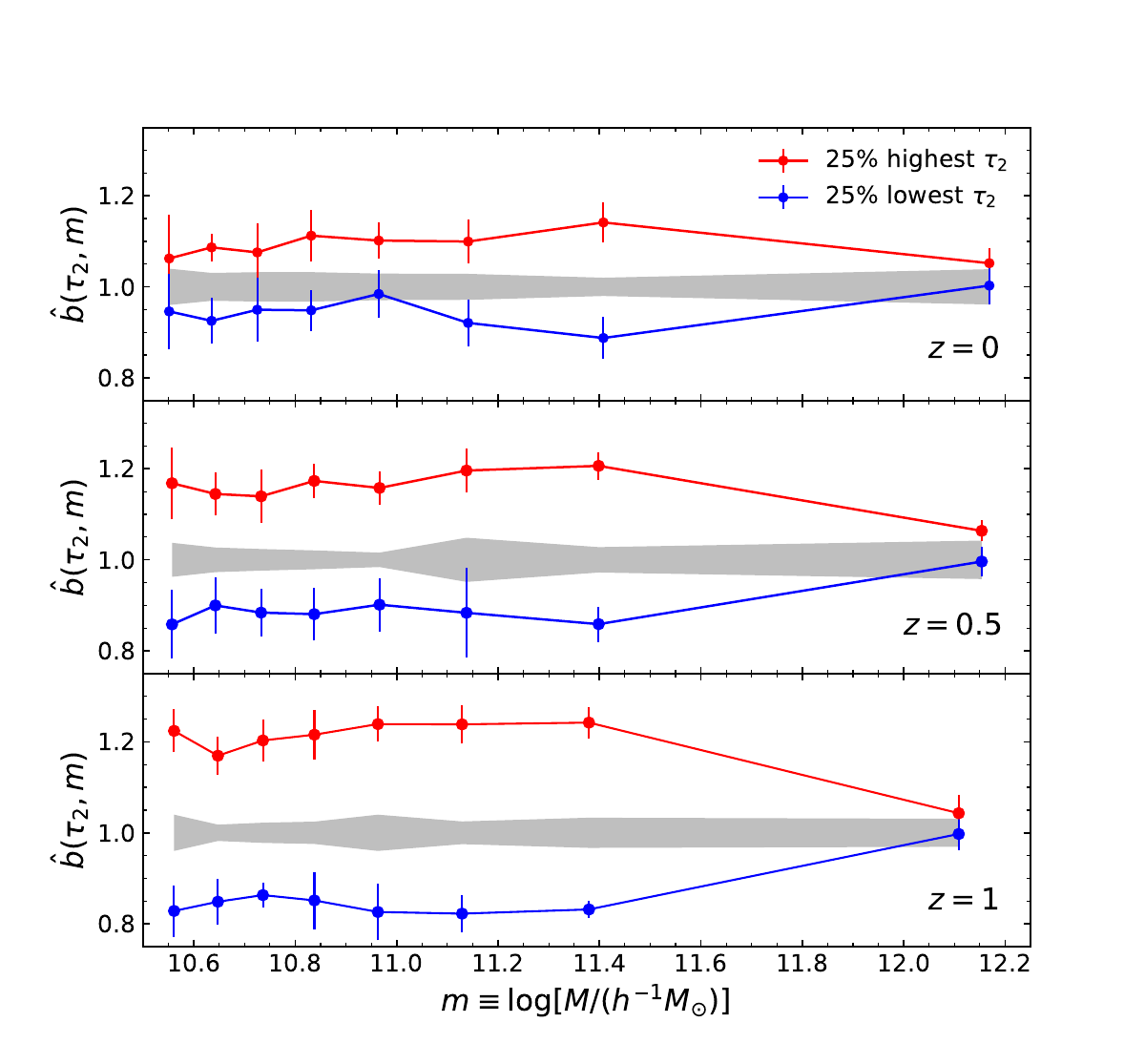}
\caption{\label{fig:tau2_bias} Same as figure~\ref{fig:tau1_bias} but for the case of $\tau_{2}$ that quantifies how weakly the triaxial shapes 
of protohalos  are aligned with the principal axes of the initial tidal field smoothed on the larger scale, $R_{f}/(h^{-1}{\rm Mpc})=2$. 
Note a similarity between the relative $\tau_{2}$ and $a_{f}$ bias (see figure~\ref{fig:age_bias}) at $z=1$.}
\end{figure}
\begin{figure}[tbp]
\centering 
\includegraphics[height=410 pt,width=395 pt]{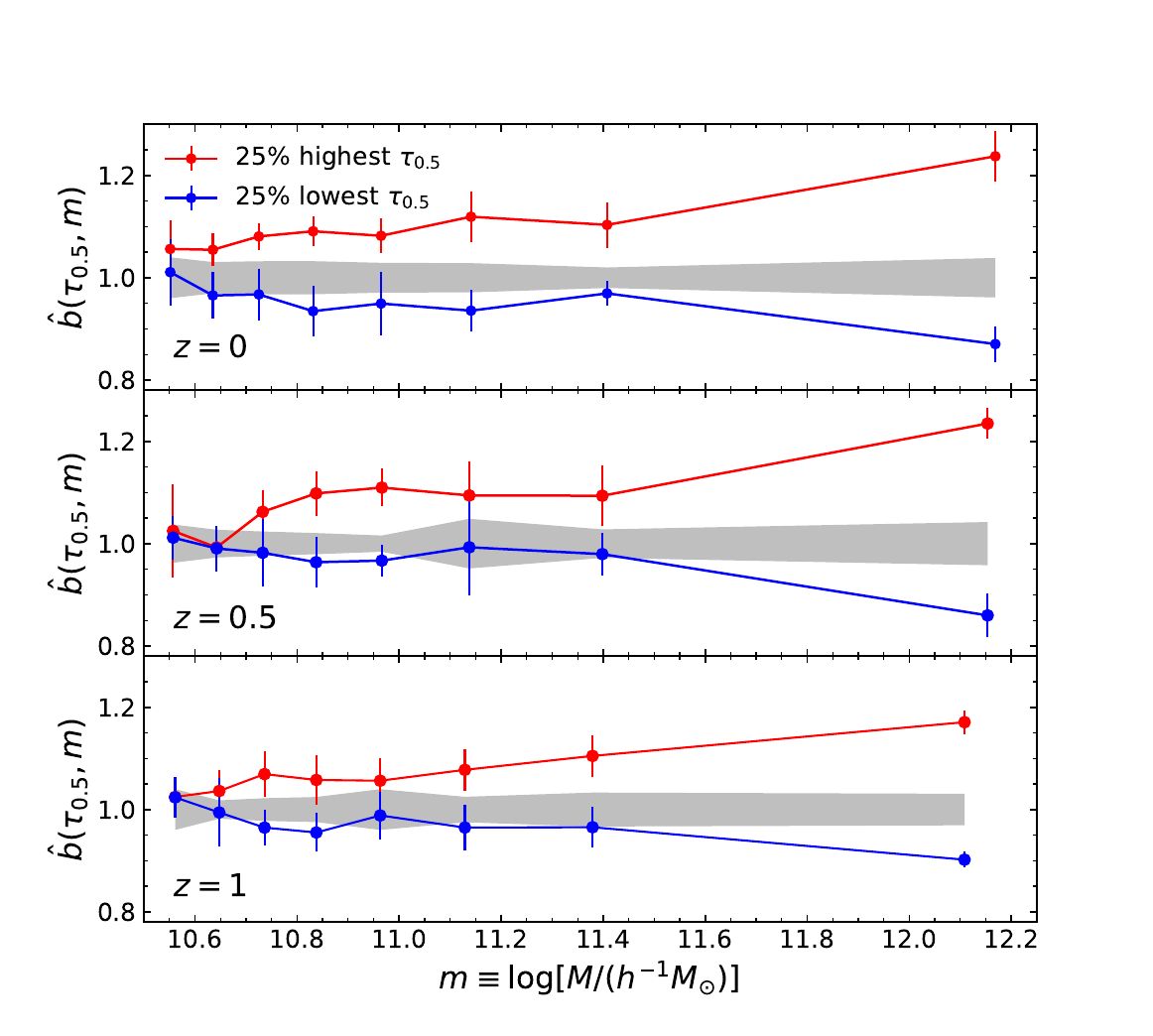}
\caption{\label{fig:tau0.5_bias} Same as figure~\ref{fig:tau1_bias} but for the case of $\tau_{0.5}$ that quantifies how weakly the triaxial shapes 
of protohalos are aligned with the principal axes of the initial tidal field smoothed on the smaller scale, $R_{f}/(h^{-1}{\rm Mpc})=0.5$.}
\end{figure}

Figure~\ref{fig:tau1_bias} shows the relative $\tau_{1}$ bias, revealing how $\hat{b}(\tau_{1},m)$ behaves as $m$ changes at different redshifts. 
The factor by which $\hat{b}(\tau_{1},m\gtrsim 12)$ enhances as $\tau_{1}$ changes from the bottom to top quartile reaches up to $\sim 1.4$ at $z=0$ 
but slightly diminishes down to $\sim 1.3$ at $z=1$.   
In the lower mass range of $m\le 11.5$, the relative $\tau_{1}$ bias has insignificantly low values at $z=0$ but tends to become significant as $z$ increases. 
The overall trend is that the $\tau_{1}$ bias mildly increases as $m$ increases at all of the three redshifts. 
Note that the behavior and amplitude of $\hat{b}(\tau_{1},m)$ at $z=1$ is quite similar to that of $\hat{b}(\lambda,m)$ shown in figure~\ref{fig:spin_bias}. 
But, at $z=0.5$ and $0$, a notable difference exists between $\hat{b}(\tau_{1},m)$ and $\hat{b}(\lambda,m)$ especially in the low mass range $m\le 11.5$. 
Unlike $\hat{b}(\lambda,m)$, no inversion to the negative bias at $m\simeq 11.5$ is exhibited by $\hat{b}(\tau,m)$ at $z=0$.

As mentioned in section~\ref{sec:intro}, however, it was already proved by ref.~\cite{splash_spin} that the inversion of $\hat{b}(\lambda,m\simeq 11.5)$ 
at $z\simeq 0$ was entirely due to the locations of abundant backsplash halos in the vicinity of massive former hosts. When the backsplash halos were 
properly excluded from the calculation of $\hat{b}(\lambda,m)$ at $z=0$, the inversion signal disappeared even at $z=0$. 
What ref.~\cite{splash_spin} also found that at $z\ge 1$ the effect of backsplash halos on $\hat{b}(\lambda,m)$ was negligible, since 
the abundance of backsplash halos decreases sharply with the increment of $z$.
Comparing $\hat{b}(\tau_{1},m)$ with the results of ref.~\cite{splash_spin}, i.e., $\hat{b}(\lambda,m)$ measured after the elimination of backsplash halos 
(see figure 3 in ref.~\cite{splash_spin}), we note that a robust similarity between the spin and $\tau_{1}$ bias factors exists even at $z=0$. 

To gain a better insight on the dependence of halo bias on $\tau_{1}$ at each redshift, we also compute $\hat{b}(\tau_{1})$ separately from the four $m$-selected 
subsamples by following the same procedure used to measure $\hat{b}(\lambda)$ in section~\ref{sec:review}. 
Figure~\ref{fig:tau1_bias2} shows the same as figure \ref{fig:spin_bias2} but with $\tau_{1}$ substituting for $\lambda$. The comparison with figure~\ref{fig:spin_bias2} 
reveals that at $z=0.5$ and $1$, the overall trends of $\hat{b}(\tau_{1})$ and $\hat{b}(\lambda)$ are indeed quite similar to each other not only in the top and bottom 
quartiles but also in the whole ranges of $\tau_{1}$ and $\lambda$. Note that even at $z=0$ where the effects of backsplash halos are dominant the two bias factors show 
similar behaviors in the range of $12\le m\le 13$. 

To see how the relative $\tau$ bias changes with the smoothing scale $r_{f}$, we repeat the whole analysis but for the case of $r_{f}=2$ and $0.5$.
Figure \ref{fig:tau2_bias} plots the same as figure \ref{fig:tau1_bias} but for the case of $r_{f}=2$, revealing a striking difference in the behavior of $\hat{b}(\tau_{2},m)$ 
from that of $\hat{b}(\tau_{1},m)$.  In contrast to $\hat{b}(\tau_{1},m)$, a mild decrease of $\hat{b}(\tau_{2},m)$ with the increment of $m$ is found in the whole mass range 
considered. Note that at $z=1$ the overall trend of the relative $\tau_{2}$ bias is quite similar to the relative age bias shown in figure~\ref{fig:age_bias}. 
Recalling the results of~\cite{MK20} that the three nonlinear mechanisms had an effect of deviating the halo age bias from the primordial trend 
by augmenting it especially in the low-mass range ($m\lesssim 12.5$), we interpret the similarity between $\hat{b}(\tau_{2},m)$ and 
$\hat{b}(a_{f},m)$ at $z=1$ as an evidence supporting the primordial origin of the halo age bias~\cite{MK20,wan-etal21}. 
We also take the low $\tau_{2}$ bias signal at $z=0$ as a result indicating that the three nonlinear processes undermines the $\tau_{2}$ bias at low redshift, 
deviating further $\hat{b}(\tau_{2},m)$ from $\hat{b}(a_{f},m)$. In other words, while the halos undergo the three nonlinear processes, they tend to lose the 
primordially generated connections between $\tau_{2}$ and $a_{f}$. 

Figure \ref{fig:tau0.5_bias} plots the same as figure \ref{fig:tau1_bias} but for the case of $r_{f}=0.5$. Although the $\tau_{0.5}$ bias is similar in its overall trend 
to the $\tau_{1}$ counterpart, the strength of the former seems to be significant even in the low-mass range of $m\lesssim 11.5$ at $z=0$, in contrast to the latter. 
The comparison of the results shown in figures \ref{fig:age_bias}-\ref{fig:tau0.5_bias} indicates that the behavior and strength of $\tau(r_{f})$ depends quite sensitively 
on $r_{f}$ and that there is two characteristic scales, say $r_{a}\simeq 2$ and $r_{s}\simeq 1$, on which the behaviors of $\tau(r_{f}=r_{a})$ and $\tau(r_{f}=r_{s})$ bias 
factors become quite similar to those of the spin and age bias, respectively, at $z\simeq 1$. Any departure of $r_{f}$ from these characteristic scales in either direction 
cause  $\hat{b}(\tau)$ to behave differently from $\hat{b}(a_{f})$ and $\hat{b}(\lambda)$, respectively.  We speculate that the primordial effect of $\tau(r_{f})$ on $a_{f}$ 
and $\lambda$ should be the strongest on these particular scales $r_{a}$ and $r_{s}$, respectively.  Moreover, given that $r_{a}$ is larger than $r_{s}$, it can be inferred 
that $a_{f}$ and $\lambda$ are affected  by the larger and smaller scale effects of $\tau$ in the primordial epoch, respectively. 
In other words, the differences between the age and spin bias may be primordially caused by the multi-scale  effects of $\tau(r_{f})$. 
At lower redshifts $z\lesssim 0.5$ when the secondary bias factors are severely modified by the nonlinear physical processes~\cite{MK20}, however, 
the primordial $\tau(r_{f})$ effects seem to dwindle away especially in the low-mass range, no matter what $r_{f}$ is. 

 \section{Summary and conclusion}\label{sec:con}

From the TNG 300-1 galactic halos in the mass range of $10.5< m\equiv \log[M/(h^{-1}M_{\odot})]\le 13$ at $z=0,\ 0.5$ and $1$, we have detected the significant 
dependence of halo bias on the primordial factors, $\tau_{0.5}$, $\tau_{1}$ and $\tau_{2}$, which denote the degree of misalignments of the triaxial protohalo shapes with 
the initial tidal tensors smoothed on the scales of $r\equiv R_{f}/(h^{-1}{\rm Mpc})=0.5,\ 1$ and $2$, respectively (see figures~\ref{fig:tau1_bias}-\ref{fig:tau0.5_bias}). 
Comparing the relative bias of the three primordial factors, $\hat{b}(\tau_{0,5},m)$, $\hat{b}(\tau_{1},m)$, and $\hat{b}(\tau_{2},m)$, with the relative age and spin bias, 
$\hat{b}(a_{f},m)$ and $\hat{b}(\lambda,m)$ (see figures~\ref{fig:age_bias}-\ref{fig:spin_bias}), we have found that at $z=1$, the overall trend of $\hat{b}(\tau_{1},m)$, is 
very similar to that of $\hat{b}(\lambda,m)$ in the entire mass range considered. 
Although an apparent difference has been found between $\hat{b}(\tau_{1},m)$ and $\hat{b}(\lambda,m)$ in the low-mass range ($m\le 11.5$) at $z=0$ and $0.5$, 
it has been realized that the exclusion of backsplash halos fully recovers the robust similarity between them~\cite{splash_spin} even in the low-mass range of $m\le 11.5$ 
at all of the three redshifts. 

It has also been found that $\hat{b}(\tau_{2},m)$ and $\hat{b}(a_{f},m)$ exhibit similar behaviors at $z=1$, but evolve in opposite directions at lower redshifts: 
as $z$ decreases, the former becoming weaker while the latter stronger.   
We have ascribed the deviation of $\hat{b}(a_{f},m)$ from $\hat{b}(\tau_{2},m)$ at $z\le 0.5$ to the three physical processes that ref.~\cite{MK20} already showed 
were mainly responsible for the age bias of low-mass halos at $z=0$: the large-scale web environments, shock heats generated during the formation of large scale structures, 
and clustering of the backsplash halos. These nonlinear processes seem to have two prolonged effects: diminishing the dependence of halo bias on the primordial factor, 
$\tau_{2}$, and enhancing the age bias of low-mass halos at lower redshifts. 

It is interesting to see that only through the calculation of bias factors is found the difference between $a_{f}$ and $\lambda$ in the scale dependence of their correlations with 
$\tau$. In our prior work~\cite{ML24} where the respective correlations of these two properties with $\tau$ were directly calculated on three different scales, the scatters around 
the mean values made it almost impossible to see whether or not their correlation strengths exhibit these variations with $r_{f}$. In other words, the similarities of $\hat{b}(\tau_{2})$ and 
$\hat{b}(\tau_{1})$ to $\hat{b}(\lambda)$ and $\hat{b}(a_{f})$, respectively, has led to this new discovery that the two halo properties, $\lambda$ and $a_{f}$, are dominantly 
affected by the primordial spin factors defined on different scales. 

We interpret $\tau(r_{f})$ as the protohalo response to the multi-scale influences of the initial tidal field, ${\bf T}(r_{f},z_{i})$.  A higher (lower) value of $\tau(r_{f})$ corresponds 
to a slower (faster) response of a protohalo to ${\bf T}(r_{f},z_{i})$.  If protohalos of same mass $M$ respond more slowly (rapidly) to ${\bf T}(r_{f},z_{i})$, then they would be 
less (more) susceptible to its effect, which will eventually produce a difference between them in their secondary properties linked with ${\bf T}(r_{f},z_{i})$. 
For the case of the spin parameter,  it is produced under the primary influence of ${\bf T}(r_{f}\simeq r_{\rm L},z_{i})$~\cite{whi84} where $r_{L}\equiv \left[4\,M/(3\pi\,\bar{\rho})\right]^{1/3}$ 
is the Lagrangian virial radius of the protohalo~\cite{wan-etal21}. Any difference in the response to ${\bf T}(r_{f}\simeq r_{\rm L},z_{i})$ would cause a difference in $\lambda$ 
among the protohalos of same mass, originating the halo spin bias and its similarity to the $\tau(r_{f}\simeq r_{L})$ bias until the effects of backsplash halos modify the 
spin bias in the lower-mass range of $m\le 11.5$ at low redshifts.  
Whereas, for the case of the formation epochs, it is determined largely by ${\bf T}(r_{f}>r_{\rm L},z)$ (i.e., the initial tidal effects on the scales larger than $r_{L}$). 
If protohalos of same mass respond more slowly (rapidly) to ${\bf T}(r_{f}>r_{\rm L},z)$ due to weak (strong) alignments of their shapes with ${\bf T}(r_{f}>r_{\rm L},z)$, 
then their formations will be delayed (expedited), which in turn will create the age bias similar to the $\tau(r_{f}> r_{L})$ bias. 
The memory for the primordial $\tau(r_{f}> r_{L})$ factor will be retained in the age bias until it is severely undermined by the nonlinear effects found by ref.~\cite{MK20}. 

Backing up the conjecture previously suggested by  refs.~\cite{MK20,wan-etal21} that the secondary halo bias must have a primordial origin regardless 
of halo mass, the current work has made two new contributions.  Firstly, we single out the primordial factor, $\tau$, the bias of which behaves like the age and 
spin bias on the galactic mass scale at $z\simeq 1$. Secondly, we have provided an explanation for the difference in behavior and tendency between the spin and 
age bias of galactic halos in terms of the primordial $\tau$ factors defined on different scales.  Our future work is in the direction of deriving an analytic model for the 
halo $\tau$ bias in terms of linear quantities from the first principle, and exploring the $\tau$ bias in a much wider range of mass.

 \acknowledgments

The IllustrisTNG simulations were undertaken with compute time awarded by the Gauss Centre for Supercomputing (GCS) 
under GCS Large-Scale Projects GCS-ILLU and GCS-DWAR on the GCS share of the supercomputer Hazel Hen at the High 
Performance Computing Center Stuttgart (HLRS), as well as on the machines of the Max Planck Computing and Data Facility 
(MPCDF) in Garching, Germany.  JL acknowledges the support by Basic Science 
Research Program through the NRF of Korea funded by the Ministry of Education (No.2019R1A2C1083855). 
JSM acknowledges the support by the National Research Foundation (NRF) of Korea grant 
funded by the Korean government (MEST) (No. 2019R1A6A1A10073437). 

\end{document}